\documentclass[conference]{IEEEtran}

\usepackage{siunitx}
\usepackage{graphicx}
\usepackage{amsmath}
\usepackage{gensymb}
\usepackage[mathcal]{eucal}
\usepackage{commath}
\usepackage{balance}

\newcommand{\tensor}[1]{\boldsymbol{\mathcal{#1}}}
\begin{document}

\IEEEoverridecommandlockouts

\title{60 GHz Blockage Study Using Phased Arrays}

\author{\IEEEauthorblockN{Christopher Slezak, Aditya Dhananjay, and Sundeep Rangan}
\IEEEauthorblockA{NYU Tandon School of Engineering\\
chris.slezak@nyu.edu, aditya@courant.nyu.edu, srangan@nyu.edu} \thanks{This work was supported in part by NSF grants 1302336, 1564142, and 1547332, NSA, NIST, and the affiliate members of NYU WIRELESS.}
}

\maketitle

\begin{abstract}
The millimeter wave (mmWave) frequencies offer the potential for enormous capacity wireless
links. However, designing robust communication systems at these frequencies requires that we understand
the channel dynamics over both time and space: 
mmWave signals are extremely vulnerable to blocking and the channel can thus rapidly appear and disappear with small movement of obstacles and reflectors. In rich scattering environments, different paths may experience different blocking
trajectories and understanding these multi-path blocking dynamics 
is essential for developing and assessing beamforming and beam-tracking algorithms.
This paper presents the design and experimental results of a novel
measurement system which uses phased arrays to perform mmWave dynamic channel measurements.
Specifically, human blockage and its effects across multiple paths are investigated with only several microseconds between successive measurements. From these measurements we develop a modeling technique which uses low-rank tensor factorization
to separate the available paths so that their joint statistics can be understood.
\end{abstract}

\section{Introduction}
Millimeter wave (mmWave) frequencies (roughly above 10 GHz) have shown tremendous promise as an enabling technology for 5G \cite{ItWillWork,rangan2014millimeter}. 
However, unlike conventional cellular communications below 6~GHz, mmWave 
signals are 
highly susceptible to blockage from objects present in the environment such as metal shelves, brick walls, and even people.
Hence, small movements of these objects can cause rapid variations in received power.
These rapid variations can complicate the operation of the entire protocol stack, and designers of 5G systems will need models which accurately characterize these effects.
Numerous procedures including channel estimation, channel quality tracking,
beamforming, rate adaptation, handover and even congestion control
all depend critically on the precise dynamics of these channels \cite{transportPerformance,channelDynamics}.

With the growing interest in mmWave, 
a number of studies have been performed to understand these dynamics better.
These studies include 
models based on ray tracing \cite{rayTracing,eliasi2015stochastic}, 
SISO measurements with fixed antennas \cite{humanActivity60,georgeBlockage}, measurements from MIMO systems with a small number of antennas \cite{MIMOmeasurements}, and measurements using omnidirectional antennas \cite{omni}.
A limitation of these previous measurement-based results is that they have been restricted
to looking at a limited number of directions at a time since they are based on small numbers
of fixed antennas.  
In a rich scattering environment, it is possible that when one path is blocked,
others may still be available.  Understanding the statistics over multiple
spatial paths during blocking events is thus necessary to fully assess 
path diversity algorithms.

In this paper, we present a novel measurement system that is capable of performing channel measurements over a large number of pointing angles almost simultaneously. These measurements allow us to jointly examine blockage across multiple paths that are present in the channel. To aid in the modeling and interpretation of the measurement data, we propose an analysis technique which uses parallel factor analysis (PARAFAC) to identify the distinct blockage trajectories that these multiple paths experience.

%The presence of objects which reflect incident waves can be used to maintain a link during a blocking event, but finding them quickly enough poses a significant challenge. The positions of these objects relative to a user and base station can vary extremely quickly, on the order of milliseconds. As an example, a recent study~ \cite{georgeBlockage}
%demonstrated that a blocking event caused by a person walking in between a TX and RX 4 meters apart can cause more than a 20 dB drop in received power in a period of tens of milliseconds. The results above show that it is indeed possible to find reflected paths capable of sustaining a link when the dominant path is blocked, but they are not sufficient to model the joint statistics of the multiple paths that are present. 

\section{Hardware Setup}

\begin{figure}

\includegraphics[width=\linewidth]{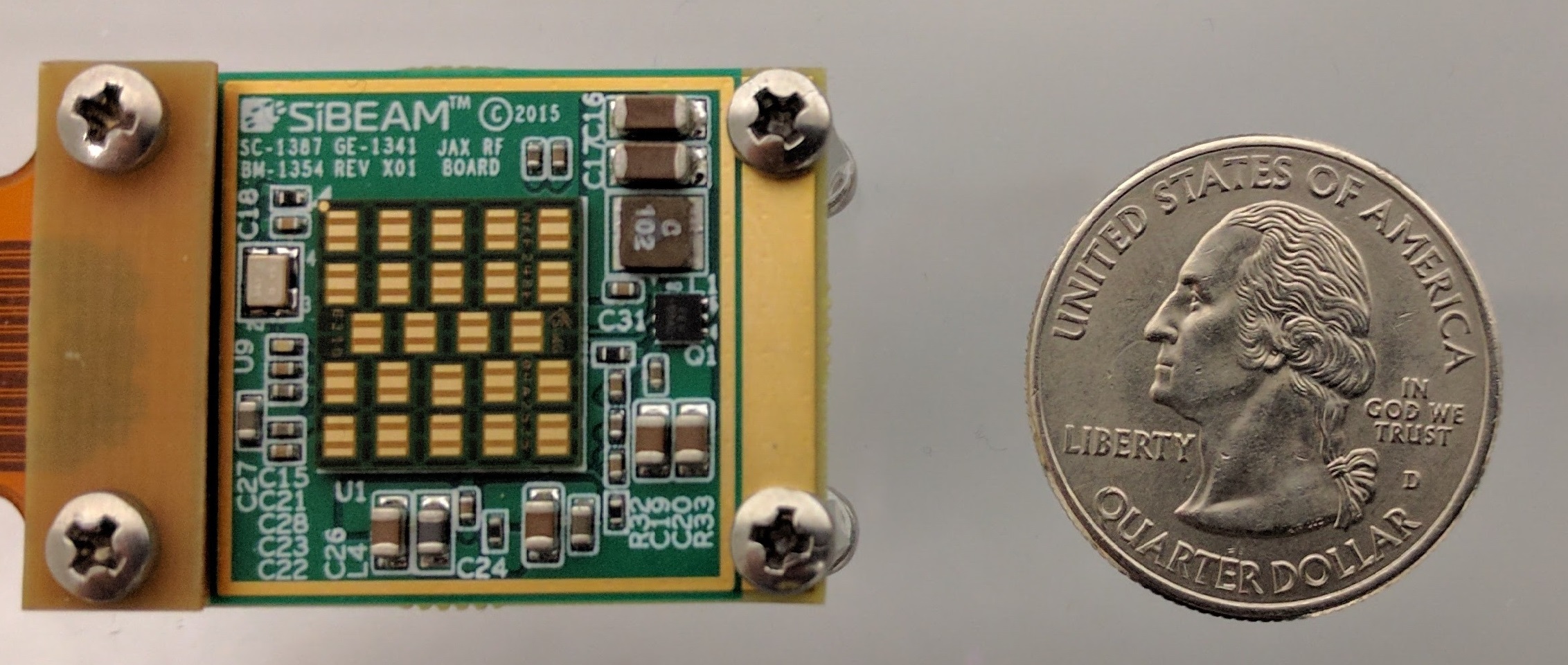}

\caption{SiBeam 60 GHz phased array. Twelve elements are active at a time.}

\label{fig:array}

\end{figure}

The critical pieces of hardware enabling this system to be built are two SiBeam 60 GHz phased arrays shown in Fig. \ref{fig:array}. These arrays have 12 steerable elements active when transmitting or receiving, and allow the user to select an arbitrary steering vector which defines the direction of arrival/departure of the array.
The antennas are designed for a narrow fixed vertical beam of approximately $\pm 10\degree$
and steerable horizontal range of approximately $\pm 45\degree$.
The total antenna gain is 23~dBi.
Applying a new steering vector to the array can be done in just a few microseconds which allows for measurements of the channel to be performed across different combinations of transmitter (TX) and receiver (RX) pointing angles with very little time between measurements. 

\begin{figure}

\includegraphics[width=\linewidth]{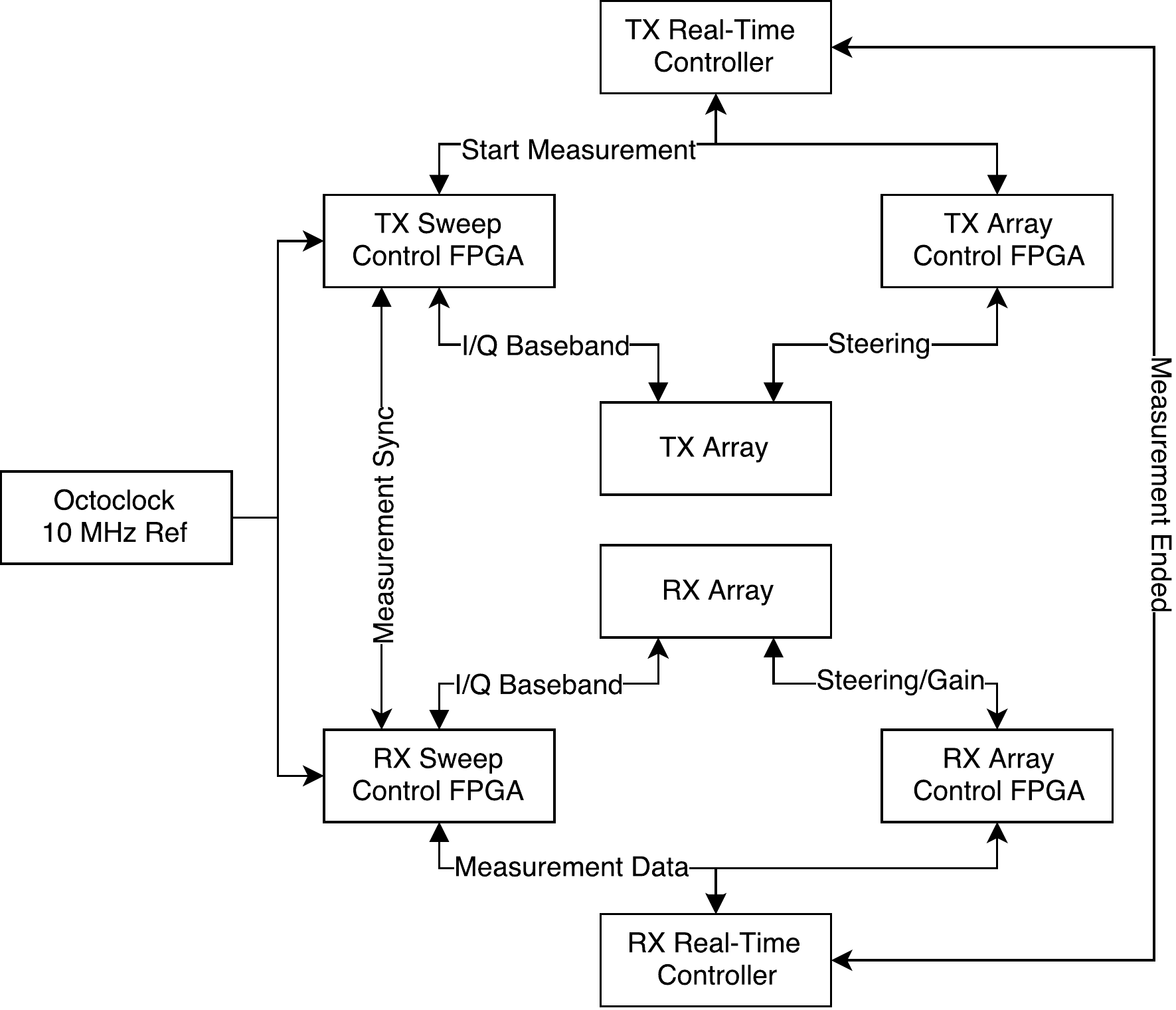}
{}
\caption{Block diagram of the measurement system.}

\label{fig:systemDiagram}

\end{figure}

These arrays are combined with a baseband system implemented using two (one each at the TX and RX) National Instruments PCI eXtenstions for Instrumentation (PXIe) chassis. Each chassis has multiple Field Programmable Gate Arrays (FPGAs), Input/Output (I/O) modules, and a host computer running a real-time operating system. Fig. \ref{fig:systemDiagram} shows a block diagram of the hardware configuration for the TX and RX. Timing synchronization is done using a common reference that is connected to both the TX and RX via a cable. Separate cables running between the two chassis allow them to send and receive control signals during the course of the measurement. These cables limit the total separation that can be achieved between the TX and RX, but for indoor measurements this is not critical.

A wideband sequence occupying 1 GHz of bandwidth at radio frequency (RF) is continuously generated at the TX chassis and sent to the TX array. The large bandwidth of this sequence gives a timing resolution of one nanosecond, crucial for indoor measurements where reflected copies of the transmitted sequence may be arriving with very small delays. No automatic gain control is implemented at the RX, so amplifier gain remains fixed throughout the measurement. This limits the dynamic range to 40 dB.

\begin{figure}

\includegraphics[width=\linewidth]{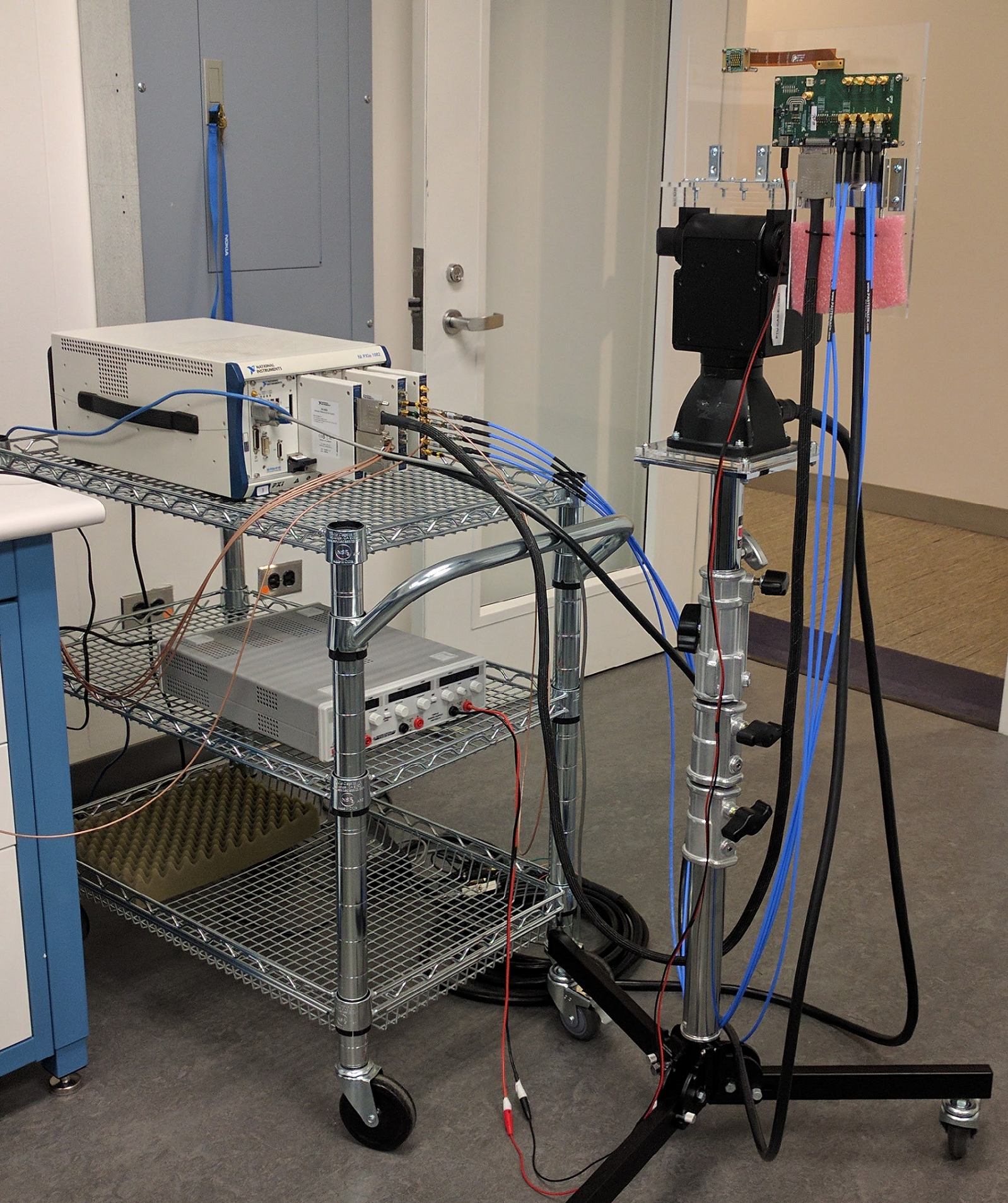}

\caption{TX baseband and RF hardware.}

\label{fig:systemPhoto}

\end{figure}

\section{Measurement Procedure}
\label{sec:procedure}

At the beginning of a measurement, the TX applies one steering vector to its array, and the RX will cycle through its codebook containing 12 predefined steering vectors, acquiring a channel impulse response for each one. The TX will then advance to the next steering vector in its codebook, and the RX will again acquire a channel impulse response for each of its 12 steering vectors. Following this process, the TX and RX will measure the channel across 144 pointing angle combinations in less than one millisecond. This 144 pointing angle scan is repeated 1750 times with one scan performed every 3 ms. Thus, a single measurement has a duration of slightly more than 5 seconds and consists of almost 5 million individual data points. There is a buffer period of 40 seconds as the data from this measurement is copied from FPGA memory to the hard disk of the chassis controller, and then the entire 5 second measurement process is repeated for several minutes until the desired number of measurements have been taken.

%\begin{figure}

%\includegraphics[width=\linewidth]{figures/room.png}

%\caption{TEMPORARY FIGURE, WILL REPLACE}

%\label{fig:room}

%\end{figure}

\begin{table}
\centering
\caption{Summary of Measurement Scenarios}
\label{tab:scenarios}
\begin{tabular}{|c | c | c | c|}
\hline
Scenario & Num. Blockers & Rotation & Num. Measurements \\ \hline
0 & 0 & No & 4 \\ \hline
1 & 1 & No & 11 \\ \hline
2 & 3 & No & 10 \\ \hline
3 & 0 & Yes & 5 \\ \hline
4 & 2 & Yes & 5 \\ \hline

\end{tabular}

\end{table}

The measurements were performed in a typical laboratory environment. Furniture included metal and wooden workbenches, office chairs, and metal shelving. A large metallic reflector was also present in the environment to ensure sufficient path diversity when validating the measurement system and associated analysis techniques. Blockage was studied under four different scenarios. In the first, a single person followed a random path through the room at a typical walking speed. In the second, three people walked through the lab following similarly random paths. In the third scenario, the TX array was rotated over a range of $\pm 45 \degree$ in the azimuth plane while there were no blockers present in the room. The angular velocity was not precisely controlled but care was taken to move approximately $90 \degree$ over the course of a single measurement. For the last scenario, the TX was rotated in the same way as before while two human blockers walked randomly throughout the room. This represents a very challenging situation to perform beam tracking/search. Measurements were also taken with no blockers present in order to understand the propagation environment of the room. Table \ref{tab:scenarios} summarizes the four scenarios and lists the number of measurements done for each. 

\section{Analysis Techniques and Results}
\subsection{Visualizing Blocking Events}
The output of a single measurement can be thought of as a three-way tensor $\tensor{X}$ where the element $x_{\tau jk}$ represents delay $\tau$ of the complex channel impulse response (CIR) measured for pointing angle combination $j$ during scan $k$. As described in Section \ref{sec:procedure} each CIR is $N_{\rm dly}=192$ samples long, there are $N_{\rm dir}=144$ point angle combinations per scan and each scan is performed $N_{\rm scan}=1750$ times. This yields a tensor of size $N_{\rm dly} \times N_{\rm dir} \times N_{\rm scan}$. The rather large size of this tensor presents a challenge when analyzing and interpreting the results of the measurement. To visualize the data, we can integrate the power delay profile (PDP) over $\tau$ to extract the received power power at pointing angle combination $j$ and scan index $k$. This reduces $\tensor{X}$ to a $N_{\rm dir} \times N_{\rm scan}$ matrix $\mathbf{P}$ with entry $p_{jk}$ given by

\begin{equation}
\label{eq:PDPMat}
p_{jk} = \sum_{\tau = 1} ^{N_{\rm dly}} \abs{x_{\tau jk}}^2.
\end{equation}

\begin{figure}

\includegraphics[width=\linewidth]{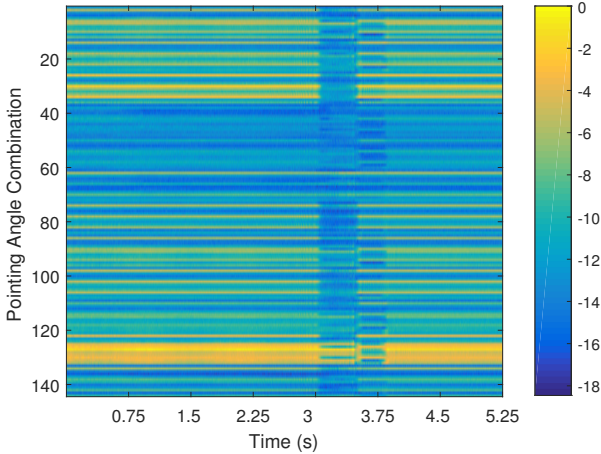}

\caption{Received power over time with two blockage events. (Scenario 1)}

\label{fig:oneWalker}

\end{figure}

Fig. \ref{fig:oneWalker} shows an example of one such matrix taken from scenario 1 where two distinct blocking events can be identified. The first blocking event begins 3 seconds into the measurement with a duration of approximately 500 ms. The second blocking event begins several milliseconds later and lasts about 350 ms. Although this representation is convenient for visual inspection of blocking events, it is not clear how this information can be used to develop a statistical model.

\begin{figure}

\includegraphics[width=\linewidth]{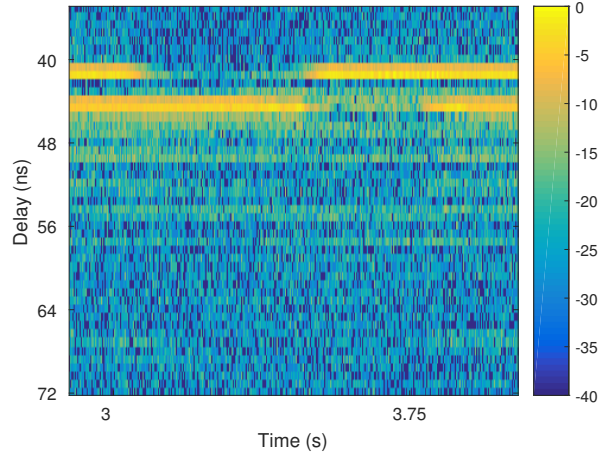}

\caption{Time Evolution of PDP for a fixed pointing angle combination.}

\label{fig:PdpTwoPath}

\end{figure}

It is apparent that a single blocking event has an observable effect on several different pointing angle combinations. This is due to the presence of sidelobes in the array pattern and the relatively large beamwidth. Further, some pointing angle combinations have contributions from multiple paths present in the channel due to these effects. For example, Fig. \ref{fig:PdpTwoPath} shows one second of PDPs acquired for a single pointing angle combination taken from the same measurement shown in Fig. \ref{fig:oneWalker}. There are two large peaks present, the earlier peak is the line-of-sight (LOS) path and the second arriving peak is a strong reflected path. This highlights the fact that is not possible to select a single pointing angle combination, examine its received power over time and conclude that it describes blockage across a single path. Instead we propose a technique which decomposes the tensor $\tensor{X}$ into a low-rank representation that identifies blockage on the dominant paths in the channel.

\subsection{Low-Rank Decomposition with Two-Way PCA}

A key observation about the measured data is that the underlying structure should be low-rank due to the small number of paths that are present in the channel \cite{lowRankEstimation}. One possible technique for identifying these dominant paths is principal component analysis (PCA). PCA is a data analysis method which has seen widespread use in a variety of fields. Mathematically, PCA is performed by taking the singular value decomposition (SVD) of a data matrix $\mathbf{X}$. The SVD of a matrix is given by

\begin{equation}
\label{eq:SVD}
\mathbf{X} = \mathbf{USV}^*,
\end{equation}
where $\mathbf{U}$ and $\mathbf{V}$ are unitary matrices and $\mathbf{S}$ is a diagonal matrix whose diagonal entries are nonnegative quantities called the singular values of $\mathbf{X}$ \cite{pcaTutorial}. Existence of the SVD is guaranteed and it is easily computed using standard linear algebra software packages. Use of this technique for our data is complicated by the fact that it is only defined for matrices but in this case the measurement data is in the form of a three-way tensor $\tensor{X}$. In this case we perform PCA by interleaving all CIRs from a single scan together. This transforms $\tensor{X}$ from a $N_{\rm dly} \times N_{\rm dir} \times N_{\rm scan}$ three-way tensor to a $N_{\rm dly}N_{\rm dir} \times N_{\rm scan}$ matrix $\mathbf{X}$ (or equivalently a two-way tensor). 
%We can further reduce the size of this matrix by discarding portions of the PDP whose delay is longer than any path we would expect to see in the environment the measurements were taken. This resulted in a final matrix of size 8784x1750.

\begin{figure}

\includegraphics[width=\linewidth]{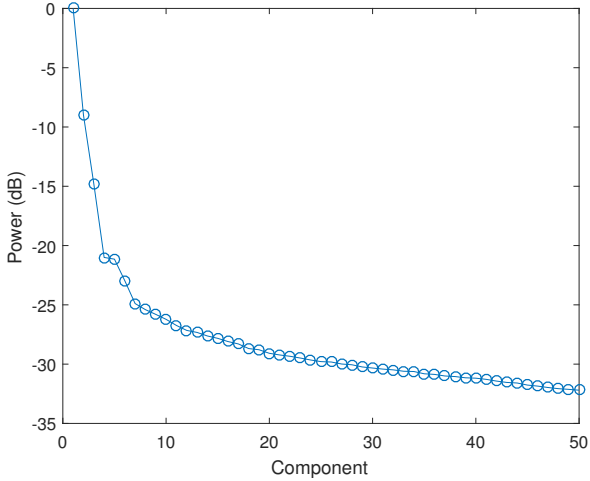}

\caption{Scree plot of the 50 largest principal components of $\mathbf{X}$.}

\label{fig:screePlot}

\end{figure}

This analysis was performed for the same measurement shown in Fig. \ref{fig:oneWalker}. A scree plot can be obtained by plotting the relative magnitudes of the diagonal elements of $\mathbf{S}$ and is shown in Fig. \ref{fig:screePlot}. As expected due to the low-rank nature of the channel, the scree plot is dominated by a few components which capture most of the variance in $\mathbf{X}$. 

\begin{figure}

\includegraphics[width=\linewidth]{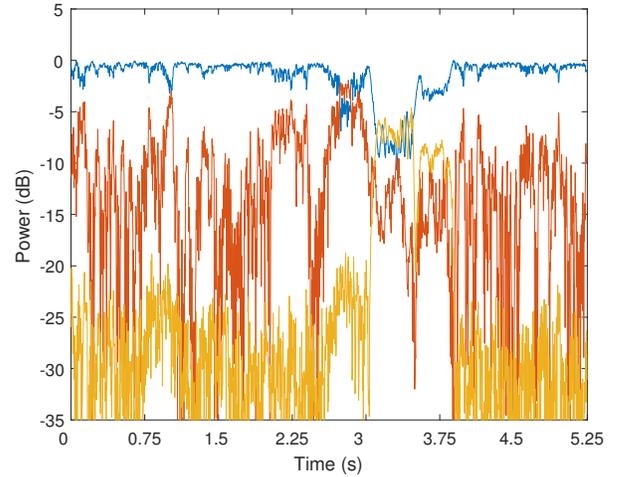}

\caption{Gain trajectories obtained from two-way PCA.}

\label{fig:twoWayPca}

\end{figure}

The gain trajectory of principal component $k$ can be obtained from this factorized form by multiplying the $k^{\rm th}$ column of $\mathbf{V}$ by the $k^{\rm th}$ element on the diagonal of $\mathbf{S}$. Fig. \ref{fig:twoWayPca} shows the gain trajectories of the three strongest principal components that were identified by this analysis, again done for the measurement shown in Fig. \ref{fig:oneWalker}. It is clear from this plot that two-way PCA failed to achieve the desired result. Component 1 is much stronger than the others throughout the duration of the measurement, and in this single component it is possible to see drops in received power which correspond to both blocking events that are visible in Fig. \ref{fig:oneWalker}. An analysis technique which can properly exploit the underlying structure of the tensor $\tensor{X}$ is needed.

\subsection{Low-Rank Tensor Decomposition with PARAFAC}
\begin{figure}

\includegraphics[width=\linewidth]{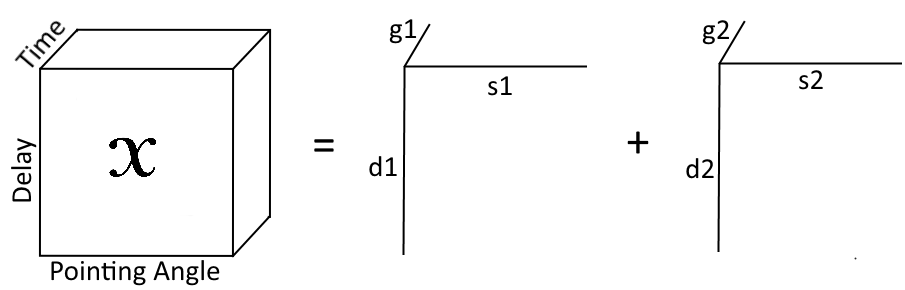}

\caption{Illustration of the PARAFAC model \cite{ParafacTutorial}.}

\label{fig:parafacModel}

\end{figure}

The failure of the more well-known PCA technique motivates the use of a less widely used but powerful method called parallel factor analysis (PARAFAC). PARAFAC is a tensor decomposition technique which can be thought of as a generalization of PCA and the SVD \cite{tensorDecompositions}. It has seen widespread use in fields such as psychometrics and chemometrics, as well as some use in the context of wireless communications \cite{ParafacChannelModel} and array processing \cite{parafacArrayProcessing}. In this particular work we wish to represent $\tensor{X}$ as a sum of rank-one tensors, with each of these rank-one tensors corresponding to a single path present in the channel. Each of these paths is described by a delay signature, spatial signature, and a gain trajectory over time. This maps very well to the PARAFAC model,

\begin{equation}
\label{eq:parafacModel}
\tensor{X} = \sum_{\ell=1} ^{L} \mathbf{d}_{\ell} \otimes \mathbf{s_{\ell}} \otimes \mathbf{g_{\ell}} + \tensor{E},
\end{equation}
where $\mathbf{d}_\ell$, $\mathbf{s}_\ell$, $\mathbf{g}_\ell$ are vectors which represent the delay signature, spatial signature, and gain trajectory of the $\ell^{\rm th}$ path, respectively, and $\tensor{E}$ is a three-way tensor representing the residual. Conceptually this is illustrated in Fig. \ref{fig:parafacModel} which shows a three-way tensor written as the sum of two rank-one tensors.

The simplicity of the PARAFAC model is a major advantage when compared to two-way PCA. Because the PARAFAC model is more restrictive than the two-way PCA model, it is able to more closely match the underlying structure of this data. In general for an $I \times J \times K$ tensor, the $L$-component PCA solution will have $L(I + JK)$ free parameters when the tensor is unraveled into a matrix suitable for PCA. The PARAFAC model on the other hand will have $L(I + J + K)$ free parameters. If this simpler model is sufficient then the extra degrees of freedom in PCA will simply lead to modeling noise or redundancies in the data \cite{ParafacTutorial}. One drawback of PARAFAC is the fact that it is computationally more involved than two-way PCA. Computing the solution requires solving an optimization problem, typically using alternating least squares. For this work the implementation of PARAFAC described in \cite{nWayToolbox} was downloaded from the MATLAB File Exchange and used to perform the decomposition. 

\begin{figure}

\includegraphics[width=\linewidth]{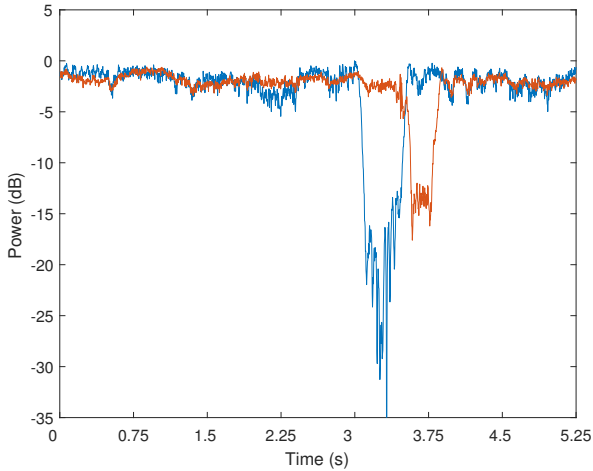}

\caption{Gain trajectories obtained from PARAFAC.}

\label{fig:parafacBlockage}

\end{figure}

Fig. \ref{fig:parafacBlockage} shows the gain trajectories for the PARAFAC model ($L=2$) computed for the measurement data shown in Fig. \ref{fig:oneWalker}. The PARAFAC model performed substantially better than two-way PCA, and has succeeded in extracting blockage trajectories suitable for the development of a statistical model.

\section{Conclusions and Further Work}

\begin{figure}
\centering
\includegraphics[width=\linewidth]{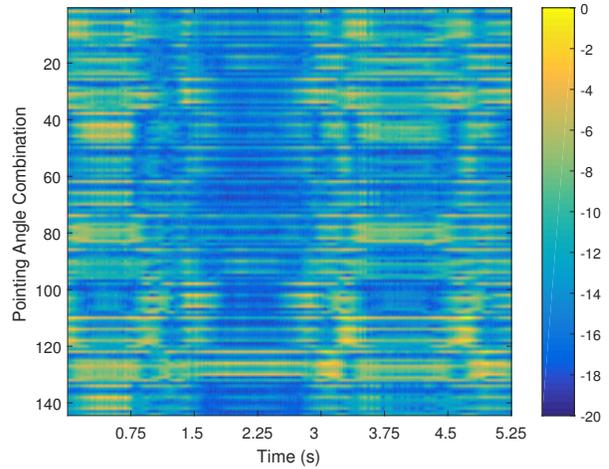}

\caption{Received power over time with human blockage and TX rotation.}

\label{fig:walkersPlusRotation}

\end{figure}

In this work we present the design of a system able to jointly measure blockage across multiple paths by rapidly and exhaustively scanning across 144 pointing angle combinations between the TX and RX. To extract the blockage trajectories from the large three-way tensors produced by these measurements, we develop an analysis technique which uses PARAFAC to construct a low-rank approximation of the measurement data. One avenue of future work is to further explore using this technique for more complicated scenarios, such as Fig. \ref{fig:walkersPlusRotation} which has blockage as well as rotation of the TX. 

Currently planned is a measurement campaign which will be done in a room that is more representative of a real-world scenario, rather than a laboratory environment as in this work. With the data from this campaign and the PARAFAC techniques described in this paper, it is possible to move on to the second stage of modeling. In this second stage, we can decompose the trajectory powers of each path into small scale and large scale components and fit a piecewise linear model to the large scale components. From this model we can assign on a per-path basis states of unblocked, entering blocking, blocked, and exiting blocking. This data can then be used to create a statistical description of 60 GHz blockage.

\bibliographystyle{IEEEtran}
\bibliography{asilomarFullPaper}

\end{document}